\documentclass[twoside]{article}
\usepackage{fleqn,espcrc2,pslatex,graphicx,cite}

\def\x#1{{x_{#1}}}
\def\nf{n_f}
\def\eps{\epsilon}
\def\e{\eps}
\def\qb{\bar q}
\def\as{\alpha_s}
\def\nn{\nonumber}
\def\R{{\rm R}}

\def\idotp#1,#2{(#1\cdot#2)}
\def\dotp#1{\expandafter\idotp#1}
\def\slash#1{{/\!\!\!\!\!#1}}
\def\osps(#1,#2,#3){\langle#1|\slash{#2}|#3\rangle}
\def\Polg{{\epsilon_g}}

\def\kq{{p_1}}
\def\kqb{{p_3}}
\def\kg{{p_2}}
\def\Li{{\rm Li}}
\def\Li(#1,#2){{{\rm Li}_{#1}(#2)}}
\def\Lii(#1,#2,#3,#4){{{\rm Li}_{#1,#2}\left(#3,#4\right)}}
\def\Liii(#1,#2,#3,#4,#5,#6){{{\rm Li}_{#1,#2,#3}\left(#4,#5,#6\right)}}
\def\z#1{{\zeta_{#1}}}
\def\RR#1{{\rm R_{#1}}}

\def\ren{\mbox{\scriptsize \rm ren}}
\def\fin{\mbox{\scriptsize \rm fin}}

\title{%
  Scattering amplitudes for e$^+$e$^- \to$ 3 jets\\
  at next-to-next-to-leading order QCD
  }

\author{%
  S. Moch\,\address{Deutsches Elektronensynchrotron DESY
    Platanenallee 6, D--15738 Zeuthen, Germany}%
  \thanks{{\tt moch@ifh.de}}, 
  P. Uwer\,\address{Institut f{\"u}r Theoretische Teilchenphysik,
    Universit{\"a}t Karlsruhe, 
    D-76128 Karlsruhe, Germany}\thanks{%
    {\tt uwer@particle.uni-karlsruhe.de}
    (speaker)},
  and S. Weinzierl\,\address{Dipartimento di Fisica, Universit\`a di Parma,
    INFN Gruppo Collegato di Parma, 43100 Parma, Italy}%
  \thanks{{\tt stefanw@fis.unipr.it}}}
\begin{document}
\begin{abstract}
\end{abstract}
\begin{abstract}
We present the calculation of the fermionic contribution to the 
QCD two-loop amplitude for $e^+e^- \to q\qb g$.
\end{abstract}
\thispagestyle{empty}
\maketitle
\section{Introduction}
Today's high energy collider experiments have reached such an accuracy
that next-to-leading order (NLO)  theoretical predictions are 
no longer sufficient. While for inclusive
observables like for example the total hadronic cross section in $e^+e^-$
annihilation the step from NLO to next-to-next-leading order (NNLO) is far
less demanding -- and has been taken a long time ago  -- the 
contrary is true for less inclusive quantities as for example jet
rates. Given these difficulties it is clear that only for specific
reactions NNLO calculations can be envisaged. Important reactions for
which one should go beyond the NLO approximation are for example: 
Bhabha scattering, $p p, p\bar p \rightarrow 2\;\mbox{jets}$, and
$ e^+e^- \rightarrow 3\; \mbox{jets}$. Among these examples the
$e^+e^-$ annihilation into 3 jets is of particular
interest. Historically this reaction was one of the first clean tests
of Quantum Chromodynamics (QCD) as the theory of strong
interaction. Today the 3-jet production in $e^+e^-$ annihilation is
still an excellent laboratory for precise tests of QCD and jet
physics. 
\begin{figure}[htbp]
  \begin{center}
    \label{fig:jetmodell}
    \includegraphics[width=7.5cm]{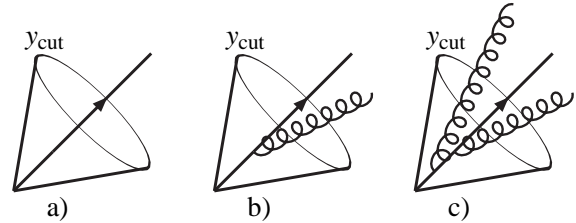}
    \caption{Modelling of a jet in leading a), next-to-leading b), and
      next-to-next-to-leading order c).}
  \end{center}
\end{figure}
If one studies how a jet is modeled in the perturbative
prediction, it also becomes obvious that only in higher order
perturbation theory will give a reliable description 
(c.f. fig. \ref{fig:jetmodell}).
Due to the fact that lepton colliders provide a very clean environment
for QCD analyses $e^+e^-\to 3\; \mbox{jets}$ is also an excellent reaction
for measureing the QCD coupling $\as$ with high accuracy. It is exactly
here where the need for NNLO predictions for 3-jet production becomes
most prominent: at present the measurements of $\as$ are plagued by
scale uncertainties of the theoretical predictions
\cite{Dissertori:2001mv} (see also H.~Stenzel, 
these proceedings). These uncertainties
arise from uncalculated higher order contributions. They can be 
reduced by a NNLO calculation.
For a 3-jet prediction at NNLO accuracy several ingredients are needed:
First the matrix elements for $e^+e^- \to q \qb ggg,\; q \qb
q'\qb'\!g$ have to be evaluated. 
They are obtained from a leading-order calculation and are
known for a long time \cite{Berends:1989yn,Hagiwara:1989pp}. 
Second the one-loop matrix elements for 
$e^+e^- \to q \qb gg, q \qb q'\qb'$ are needed. They have been
calculated in 
refs. \cite{Bern:1997ka,Bern:1998sc,Glover:1997eh,Campbell:1997tv}. 
Finally one needs the NNLO amplitudes for 
$e^+e^- \to q \qb g$ \cite{Garland:2001tf,Garland:2002ak,Moch:2002an,Moch:2002hm}. 
In the following the calculation of the NNLO matrix
elements is reported. One should keep in mind that the combination of
these 3 ingredients into an infrared finite cross section is still a 
highly non-trivial -- and so far unsolved -- problem.

\section{Old and new approaches to calculate NNLO amplitudes}
Due to the tremendous activities in the field enormous progress has
been made in the past. All the relevant double box scalar
integrals have been calculated 
\cite{Smirnov:1999gc,Tausk:1999vh,Smirnov:2000vy,Smirnov:2000ie}. 
For a few reactions these
results have been used to obtain NNLO scattering amplitudes 
\cite{Bern:2000dn,Anastasiou:2000kg,Anastasiou:2000ue,Anastasiou:2000mv,Anastasiou:2001sv,Glover:2001af,Garland:2001tf,Bern:2000ie,Bern:2001df,Bern:2002tk,Garland:2002ak,Anastasiou:2002zn}.
Today one can say that a {\it standard approach} to perform these
calculations exist. Essentially it consists of three steps. First one
generates automatically the relevant Feynman diagrams. In a second
step the reduction of tensor integrals to scalar integrals is done via
Schwinger parametrization \cite{Tarasov:1996br,Tarasov:1997kx}. 
The scalar integrals obtained in this step
appear in shifted dimensions and with raised powers of the
propagators. In a last step, equations derived from partial integration
and Lorentz invariance are used to reduce the scalar integrals to a
small set of master integrals 
\cite{'tHooft:1972fi,Chetyrkin:1981qh,Gehrmann:1999as}. 
The final result is than obtained in
terms of these master integrals. They have to be calculated
analytically. The bottleneck of this approach is that one has to
create and to solve a huge system of equations to obtain the desired reduction 
to master integrals. In contrast to the naive expectation, topologies
which at first glance look not so difficult may involve much more 
work than the more complicated once like for example the double box.
In ref. \cite{Moch:2001zr} we have proposed a different method. 
The basic idea is that
one applies only obvious reductions like for example 
the {\it triangle rule}. In particular one does not perform a complete
reduction to master integrals. Instead one calculates the scalar
integrals in higher dimensions and with raised powers of the
propagators directly. 
\begin{figure}[htbp]
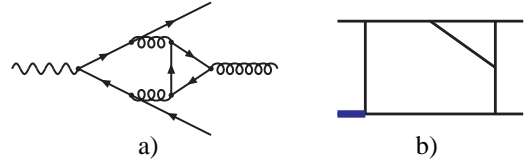

  \begin{center}
    \parbox{3.7cm}{
      \begin{center}
        \includegraphics[scale=0.5]{figs.001}
      \end{center}
      }
    \parbox{3.7cm}{
      \begin{center}
        \includegraphics[scale=0.7]{figs.002}
      \end{center}
      }\vspace*{-0.4cm}
    \parbox[t]{3.7cm}{\centerline{a)}}\parbox[t]{3.7cm}{\centerline{b)}}
    \caption{Tensor reduction of the general pentabox topology a) to
      scalar integrals in higher dimensions and raised powers of the
      propagators b).}
    \label{fig:pentabox-a}
  \end{center}
\end{figure}
To illustrate the method let us consider the so-called penta-box. 
A corresponding Feynman diagram is shown in fig. \ref{fig:pentabox-a}a. 
Applying Schwinger parametrization one arrives at the scalar topology
shown schematically in  fig. \ref{fig:pentabox-a}b. An obvious
application of the triangle rule allows the elimination of two
propagators \cite{Anastasiou:1999bn}. The resulting topology is often called C-topology.
This reduction is shown in fig. \ref{fig:ctopo}. As mentioned earlier
the C-topology appears in higher dimensions $(d=2m-2\e,\; m\in N)$ and with 
raised powers of the propagators $(\nu_1,\ldots,\nu_5)$.
\begin{figure}[htbp]
  \begin{center}
    \includegraphics[width=7.3cm]{figs.003}
    \caption{Application of the triangle rule yielding the C-topology
      $\mbox{C}(m={d\over 2}+\e,\nu_1,\nu_2,\nu_3,\nu_4,\nu_5)$}
    \label{fig:ctopo}
  \end{center}
\end{figure}
In the approach advocated here an analytic expression for the
C-topology is needed. Such an expression can be obtained in terms
of infinite sums \cite{Moch:2001zr}. The generic structure is shown in
eq.~(\ref{result_Ctopo}).
\begin{eqnarray}
  \label{result_Ctopo}
  &&\mbox{C}(m={d\over 2}+\e,\nu_1,\ldots,\nu_5)  \sim 
 \sum\limits_{i_1=0}^\infty
 \sum\limits_{i_2=0}^\infty
 \frac{x_1^{i_1}}{i_1!}
 \frac{x_2^{i_2}}{i_2!}\bigg[
 \nonumber \\&&\hspace{-0.9cm}
  x_1^{2m-2\eps-\nu_{1235}} x_2^{2m-2\eps-\nu_{2345}}
{\Gamma(i_1+2m-2\eps-\nu_{125})\over \Gamma(i_1+1+2m-2\eps-\nu_{1235})}
 \nonumber \\&&\hspace{-0.7cm}
   \times\frac{
     \Gamma(i_1+i_2+3m-3\eps-\nu_{12345})}
   { \Gamma(i_2+1+2m-2\eps-\nu_{2345})} 
 \nonumber \\
 & & \hspace{-0.7cm}
 \times
 \frac{
   \Gamma(i_1+i_2+2m-2\eps-\nu_{235})
   \Gamma(i_2+2m-2\eps-\nu_{345}) 
   }{\Gamma(i_1+i_2+4m-4\eps-\nu_{12345}-\nu_5)}
\nonumber \\
 & &
+ \ldots\bigg] 
\end{eqnarray}
It is obvious that such a representation is completely useless unless
one is able to evaluate the $\eps$-expansion in terms of known
functions. Fortunately this is possible using a few basic algorithms
for {\it nested sums} \cite{Moch:2001zr}. 
The basic idea here is that one rewrites the
sums using the relation
\begin{equation} \sum\limits_{i=0}^\infty \sum\limits_{j=0}^\infty 
  a_{ij}
  =   
  a_{00}
  + \sum\limits_{j=1}^\infty a_{0j}
  + \sum\limits_{i=1}^\infty a_{i0}
  + \sum\limits_{n=1}^\infty \sum\limits_{j=1}^{n-1} a_{j(n-j)}.
  \label{eq:nestedsums1}
\end{equation}
Furthermore one can expand the $\Gamma$-functions by the means of
\begin{eqnarray}
  {\Gamma(-n+\e)} &=& 
  {(-1)^n\over \e n!}{\Gamma(1+\e)}
  ( 1 + S_1(n)\e \nn\\
  &&
  \hspace{-2cm}+ S_{1,1}(n)\e^2 
  + S_{1,1,1}(n)\e^3 +
  S_{1,1,1,1}(n)\e^4+\ldots
  \label{eq:expandG}
\end{eqnarray}
and related identities \cite{Vermaseren:1998uu}.
Here $S_{1,\ldots}(n)$ denote the harmonic sums, 
see e.g. ref. \cite{Vermaseren:1998uu}. Inspecting
the sums which appear using eq.~(\ref{eq:nestedsums1}) and 
eq.~(\ref{eq:expandG}) one observes 
that the algorithms needed are just a generalization of the algorithms
studied in ref. \cite{Vermaseren:1998uu}. 
In particular defining the nested sums 
as a generalization of the harmonic sums by 
\begin{eqnarray}
  &&\hspace{-0.7cm}S(n;m_1,...,m_k;x_1,...,x_k)  
  = \sum\limits_{i_1=1}^n\sum\limits_{i_2=1}^{i_1}
 \cdots \sum\limits_{i_k=1}^{i_{k-1}} {{x_1^{i_1}\over i_1^{m_1}}
   \cdots {x_k^{i_k}\over i_k^{m_k}}}\nn\\
  &&= \sum\limits_{i=1}^n {{\frac{x_1^i}{i^{m_1}}}}
 S({{i}};m_2,...,m_k;x_2,...,x_k)  
\end{eqnarray}
the following four basic operations are sufficient to evaluate 
the $\eps$-expansion of the C-topology in terms of multiple
polylogarithms \cite{Goncharov}:
\begin{enumerate}
\item Multiplication
  \begin{equation}
    {{S(n;m_1,...;x_1,...)\,\times\, S(n;m_1',...;x_1',...)}}
  \end{equation}
\item Sums involving {{$i$}} and {{$n-i$}}:
  \begin{eqnarray}
    &&\sum\limits_{i=1}^{n-1}
    \frac{x_1^i}{i^{m_1}} S({{i}};m_2...;x_2,...)\nn\\
    &&\times
    \frac{{x_1'}^{n-i}}{(n-i)^{m_1'}} S({{n-i}};m_2',...;x_2',...)
  \end{eqnarray}
\item Conjugations:
  \begin{equation}
    \hspace{-0.6cm}\sum\limits_{i=1}^n {{\left( \begin{array}{c} n \\ i 
            \\ \end{array} \right) (-1)^i}} 
    \frac{x_0^i}{i^{m_0}} S(i;m_1,...,m_k;x_1,...,x_k)
  \end{equation}
\item Sums involving {{binomials}} and {{$n-i$}}:
  \begin{eqnarray}
    &&\sum\limits_{i=1}^{n-1}
    {{\left( \begin{array}{c} n \\ i \\ \end{array} \right)}}
    \left(-1\right)^i
    \frac{x_1^i}{i^{m_1}} S(i;m_2...;x_2,...)
    \nn\\&&
    \times \frac{{x_1'}^{n-i}}{(n-i)^{m_1'}} S({{n-i}};m_2',...;x_2',...)
  \end{eqnarray}
\end{enumerate}
In ref. \cite{Moch:2001zr} we have given explicit algorithms for 
the four basic 
operations. These algorithms are well suited for an implementation in a 
computer algebra system. For their implementation we  used 
Form3 \cite{Vermaseren:2000nd} and Ginac 
\cite{Bauer:2000cp,Weinzierl:2002hv}.
Using these programs we calculated more than 300
C-topologies analytically. In particular we have checked that our
results for the two master integrals 
${\rm C}(2,1,1,1,1,1), {\rm C}(2,1,1,1,1,2)$  
agree with those given by Gehrmann
and Remiddi \cite{Gehrmann:2000zt}. 
All the simpler topologies can be calculated
in the same way.
\section{Results}
The amplitude ${\cal A}_\gamma$ for the process 
\begin{equation}
  e^+e^-\to \gamma^\ast\to  q(\kq) g(\kg)\qb(\kqb)
\end{equation}
can be written as a leptonic part $L_\mu$ multiplied by a
hadronic part $H_\mu$:
\begin{eqnarray}
 {\cal A}_\gamma &\sim& {1\over s} \, H_\mu\, L^\mu \nn\\
 &&\hspace{-1.1cm}=
 {1\over s} \, g_s (H_\mu^{(0)} 
 + {\as\over 2\pi} H_\mu^{(1)}
 + \left({\as\over 2\pi}\right)^2 H_\mu^{(2)}+\ldots)\, L^\mu,
\end{eqnarray}
with $s=(\kq+\kg+\kqb)^2$ and $g_s =\sqrt{4 \pi \alpha_s} $.
Furthermore  the hadronic part can be decomposed according to the
colour structure. In particular we have
\begin{eqnarray}
  H_\mu^{(2)} &=& T^a_{q\bar q} \bigg( 
   N^2 H_{2,\mu}^{(2)} + H_{0,\mu}^{(2)} + {1\over N^2} H_{-2,\mu}^{(2)}
   \nn\\&+&
   { \nf N H_{\nf,1 ,\mu}^{(2)}
     + {\nf\over N} H_{\nf,-1,\mu}}^{(2)}
   + \nf^2 H_{\nf^2,\mu}^{(2)}\nn\\&+&
   \Sigma_f \left({4\over N}-N\right) H_{\Sigma_f,\mu}^{(2)} \bigg),
\end{eqnarray}
with $T^a_{q\bar q} $ the generator of the $SU(N)$ gauge group and
$\nf$ the number of massless quarks.
The last contribution arise from a closed quark loop coupled directly
to the photon. (The factor  $\Sigma_f$ involves a sum over the charges
of the active quarks.) We note that the contribution proportional to $\nf^2$
is completly determined by the renormalization. It does not involve any
non-trivial two-loop topology.
In this talk we discuss only the contributions proportional to
$ \nf N$ and ${\nf\over N}$. The hadronic contribution can also be
decomposed according to the spinor structure:
\begin{eqnarray}
  H_\mu
  &=& c_1 \*{1\over  s}\* \osps(\kq,\kg,\kqb) \* \Polg_\mu
  \nn\\&&
  + c_2 \*{1\over s^2}  \* \osps(\kq,\kg,\kqb) \* \dotp{\Polg,\kq} \* 
  \kqb_\mu  
  \nn\\&&
  + c_3 \*{1\over s^2}  \* \osps(\kq,\kg,\kqb) \* \dotp{\Polg,\kqb} \* 
  \kq_\mu 
  \nn\\&&
  +
     c_4 \*{1\over s^2}  \* \osps(\kq,\kg,\kqb) \* \dotp{\Polg,\kq} \* 
  \kq_\mu \nn\\&&
  +\ldots
  +
  c_{13}\*{1\over s^2}  \* \osps(\kq,\kg,\kqb) \* \dotp{\Polg,\kqb} \* 
  \kg_\mu 
\end{eqnarray}
where the scalar functions $c_1-c_{13}$ depend only on the ratios 
$x_1=2(\kq\cdot \kg)/s$, $x_2=2(\kg\cdot \kqb)/s$. Due to various constraints arising from gauge invariance
or quark anti-quark antisymmetry only 4 of them are needed. All the
remaining ones can be expressed in terms of these 4 functions which 
we chose to be $c_2,\, c_4,\, c_6$ and $c_{12}$.
The perturbative expansion in $\alpha_s$ of the functions 
$c_i$ is defined through
\begin{equation}
  c_i = g_s \left(
    c_i^{(0)} \!
    +\left( \frac{\alpha_s}{2\pi}\right) c_i^{(1)} \!
    +\left( \frac{\alpha_s}{2\pi}\right)^2 c_i^{(2)}\!
    + O(\as^3) 
  \right)\!\!.
\end{equation}
Using
the methods described in the previous section we calculated all
the 13 functions and used the constraints as an additional check of
our calculation.
Furthermore we checked the structure of the ultraviolet as well as 
the infrared divergencies \cite{Catani:1998bh}. 
We have also compared our results
with results published recently \cite{Garland:2001tf,Garland:2002ak}. 
In refs. \cite{Garland:2001tf,Garland:2002ak} the result is written
in terms of 2-dimensional harmonic polylogarithms instead of the multiple
polylogarithms used here. Expressing the 2-dimensional 
harmonic polylogarithms in terms of multiple polylogarithms we found
complete agreement for the $ \nf N$ and ${\nf\over N}$ contribution
calculated here.

Following ref. \cite{Garland:2002ak} we used the general 
structure of the infrared
divergencies as described by Catani \cite{Catani:1998bh} to define
the finite parts of the coefficients $c_i$:
\begin{eqnarray}
   c_{i}^{(2),\fin} &=& c_{i}^{(1),\ren} - {\bf I}^{(1)}(\eps) c_{i}^{(1),\ren}
   - {\bf I}^{(2)}(\eps) c_{i}^{(0)} \, , 
  \label{eq:def-results2}
\end{eqnarray}
with the one- and two-loop insertion operators 
${\bf I}^{(1)}(\eps)$ and ${\bf I}^{(2)}(\eps)$ given in ref. 
\cite{Catani:1998bh}.
As an example, we present the result for the $\nf N$-contribution to 
the finite part $c_{4}^{(2),\fin}$ at two loops,
\begin{eqnarray}
  &&\hspace{-0.6cm}c_4^{(2),\fin}(\x1,\x2)\bigg|_{\nf N} =
  \nf\*N\*\bigg(
  -{19 \over 18} \*{1\over \x1\*(1-\x2)}
  \nn\\&&
  \hspace{-0.6cm}
  + {1 \over 36} \*{\ln(\x2) \over (1-\x2)}\*
  \bigg[
  36\*{(1+\x1)\over (\x1+\x2)}-{59\over \x1}-{50\over \x1\*(1-\x2)}
  \nn\\&&
  \hspace{-0.6cm}
  +12\*{(1-3\*\x1)\*(1-\x1)\over \x1\*(1-\x1-\x2)}
  \bigg]
  +{(\x1-\x2)\over (\x1+\x2)^3}\*\RR1(\x1,\x2)
    \nn\\&&
    \hspace{-0.6cm}
 -\bigg[{5 \over 12} \*{(2-\x2)\over \x1\*(1-\x2)^2}
  +{2\over (\x1+\x2)^2}
  \bigg]\*\Li(2,1-\x2)
    \nn\\&&
 +{1 \over 4} \*{(2-\x2)\over \x1\*(1-\x2)^2}\*\bigg[
 \ln(\x2)^2+{1 \over 3} \*\z2\bigg]
  \nn\\&&
    \hspace{-0.6cm}
  -{1 \over 12} \*{(7\*\x1-5\*\x2-6\*\x1\*\x2+5\*\x2^2+\x1^2)
    \over (1-\x2)\*\x1\*(1-\x1-\x2)\*(\x1+\x2)}\*\ln(\x1)
  \nn\\&&
  \hspace{-0.7cm}
  -{\R(\x1,\x2)\over (1-\x1-\x2)\*(1-\x2)\*(\x1+\x2)}\*
  \bigg[{17 \over 6 }\*\x2-{35 \over 12} \*\x1
  \nn\\&&
  \hspace{-0.5cm}
  -{19 \over 12} +{1 \over 12} \*{(1+\x1)\over (1-\x2)}
  -{1 \over 3} \*{(1-3\*\x1)\*(1-\x1)\over (1-\x1-\x2)}
  \nn\\&&
  +{5\*\x2^2 \over 12\*\x1} -{5 \over 6} \*{\x2\over \x1}
  +2\*{\x1\*(1+\x1)\over (\x1+\x2)}\bigg]
  \nn\\&&
  \hspace{-0.5cm}
  -\bigg[
 {1 \over 12} \*{(2-\x2)\over \x1\*(1-\x2)^2}
  -{2\over (\x1+\x2)^2}
  \bigg]
  \*\Li(2,1-\x1)
  \bigg)
  \nn\\&&
  \hspace{-0.6cm}
  -i\*\pi\*\nf\*N\*\bigg(
  {(2-\x2)\*\ln(\x2)\over 2\*\x1\*(1-\x2)^2} 
  +{1 \over 2} \*{1\over \x1\*(1-\x2)}\bigg) 
.\label{eq:c4nn}
\end{eqnarray}
\def\Li{{\rm Li}}%
We have introduced the function $\R(\x1,\x2)$, 
which is well known from \cite{Ellis:1981wv},
\begin{eqnarray}
  \label{eq:Rfunction}
  &&\hspace{-0.5cm}\R(\x1,\x2) =
  \bigg( {1\over 2} \ln(\x1) \ln(\x2)
    -\ln(\x1) \ln(1\!-\!\x1)
    \nn\\&&
    +{1\over 2} \z2-\Li_{2}(\x1) \bigg) +
  (\x1\leftrightarrow\x2).
\end{eqnarray}
In addition, it is convenient, to define the symmetric function
$\RR1(\x1,\x2)$, which contains a particular combination of 
multiple polylogarithms \cite{Goncharov},
\begin{eqnarray}
  \label{eq:RRfunctions}
  && \RR1(\x1,\x2) = \bigg(
  \ln(\x1) \Li_{1,1}\left({\x1 \over \x1\!+\!\x2},\x1\!+\!\x2 \right)
  \nn\\&&\hspace{-0.7cm}
  -{1 \over 2} \z2 \ln(1\!-\!\x1\!-\!\x2)
  -{1 \over 2}\ln(\x1) \ln(\x2) \ln(1\!-\!\x1\!-\!\x2) 
  \nn \\ & &
  -\ln(\x1) \Li_{2}(\x1\!+\!\x2)
  -\Li_{1,2}\left({\x1 \over \x1\!+\!\x2},\x1\!+\!\x2 \right)
  \nn \\ & &
  -\Li_{2,1}\left({\x1 \over \x1\!+\!\x2},\x1\!+\!\x2 \right)
   +\Li_{3}(\x1\!+\!\x2) \bigg)\nn\\&& + (\x1\leftrightarrow\x2)\, .
\end{eqnarray}
As one can see all multiple polylogarithms have simple arguments. 
As a consequence it is straightforward to obtain the analytic
continuation, which can be used for the crossing of the amplitude.

\section{Conclusion}
In this talk we have presented the calculation of a specific colour
structure of the QCD two-loop amplitude $e^+e^-\to q\qb g$. The
computation has been done by a completely new method developed by the
authors \cite{Moch:2001zr}. 
It has been shown that the approach which is based 
on nested sums and
their algebraic properties provides a very efficient
method for two-loop calculations. 
Furthermore the tools developed in this approach can also be
used for a systematic expansion of generalized hypergeometric
functions.
In addition the presented results provide a very important cross check
on recently obtained results   \cite{Garland:2001tf,Garland:2002ak}.\\

\parindent0cm
{\bf Acknowledgment:} P.U. would like to thank the organizers for
the pleasant atmosphere at QCD02.

\end{document}